\documentclass[aip,reprint]{revtex4-1}
\usepackage[pdftex]{graphicx,graphics}
\usepackage{amssymb,amsmath}

\begin{document}

\normalsize

\title{Passage of an ion-acoustic solitary wave through the boundary between an~electron-ion plasma and a negative ion plasma}

\author{Yu. V. Medvedev}

\affiliation{Joint Institute for High Temperatures, Russian Academy of Sciences, Moscow, 125412 Russia}

\begin{abstract}

The passage of an ion-acoustic solitary wave through the boundary between an electron-ion plasma and a negative ion plasma is considered.
After the ion-acoustic solitary wave enters the region of another plasma, a disturbance arises, from which an ion-acoustic solitary wave and a chain of oscillations form over time.
The amplitude of the ion-acoustic solitary wave after passage through the boundary changes in such a way that its value in the electron-ion plasma is greater than its value in the  negative ion plasmas.
An exception is the case of a compressive ion-acoustic solitary wave propagating through the negative ion plasma and having an amplitude exceeding the critical amplitude in the electron-ion plasma.
Such an ion-acoustic solitary wave, when entering an electron-ion plasma, releases an excess of energy to accelerate positive ions and thereby reduces its amplitude below the critical value.
The dependence of the amplitude of an ion-acoustic solitary wave established after the boundary crossing  on its initial amplitude is determined.
The passage of an ion-acoustic solitary wave through a layer of negative ion plasma surrounded by electron-ion plasmas is considered.
It is shown that the passage of a rarefactive ion-acoustic solitary wave from the negative ion plasma into the electron-ion plasma causes disturbance, in which accelerated and trapped negative ions can be observed.

\end{abstract}

\maketitle

\section{Introduction}

Ion-acoustic solitary waves can propagate both in an electron-ion plasma and in a multicomponent plasma consisting of three or more kinds of particles.
Of particular interest is a negative ion plasma, consisting of electrons and positive and negative ions.
In nature, negative ions are present in the D layer and in the lower part of the E layer of the ionosphere.
In the first experimental devices, a negative ion plasma was created using thermal ionization \cite {vGoeler} or injection of the electronegative gas $ \mathrm {SF} _6 $ into the region of an electron-ion plasma  \cite {Arnush}.
Over time, the basic properties of such plasmas were investigated and the fields of their application were established.
In particular, the possibility of using  negative deuterium ions $ \mathrm {D ^ -} $ in ion sources for neutral injection into controlled fusion devices was considered \cite{Bacal}.
A negative ion plasma turned out to be a convenient medium for experimental studies of various nonlinear phenomena, including phenomena arising from the interaction of nonlinear structures with each other.
Note that in a collisionless negative ion plasma, nonlinear structures such as ion-acoustic solitary  waves or collisionless shock waves, can exist not only as compressive waves, but also as rarefactive waves.
 
Considerable attention was paid to the study of compressive  ion-acoustic solitons and rarefactive ion-acoustic solitons, as well as  their interactions with each other.
In experiments, a negative ion plasma was created using a multi-dipole double-plasma machine \cite {Nakamura84a, Nakamura84b, Nakamura85a, Nakamura85b, Nakamura85c}, or a specially developed large multi-dipole plasma device 
\cite{Cooney91,Cooney93a,Cooney93b,Cooney93c,Cooney96}.   
When studying shock waves in a negative ion plasma, the latter was created by injecting $ \mathrm {SF} _6 $ into a single-ended Q machine \cite {DAngelo, Takeuch98}. 
A method of creating a negative ion plasma in a single-ended Q-machine, in which negative ions are formed from fullerene $\mathrm{C}_{60}$ and positive ions are $ \mathrm {K} ^ + $, has been described \cite {Sato94, Hirata94, Hirata96, OohHatak1}.
In this case the formation of negative ions occurs in a small limited region, which is surrounded by an electron-ion plasma region.
The processes occurring during such a local formation of negative ions were studied by the particle-in-cell  method \cite {OohHatak2}.
Note that in most of the above articles, as well as in a number of other articles, the propagation of ion-acoustic solitary waves and their interaction with each other were investigated analytically and numerically.
  
Laboratory methods for creating a negative ion plasma are characterized by the fact that in many cases there arises not only a region occupied by a negative ion plasma, but also a region occupied by an electron-ion plasma adjacent to it.
Therefore, in an experimental study of ion-acoustic solitary waves in a negative ion plasma, the problem of the passage of ion-acoustic solitary waves through the boundary between plasmas may arise.
In addition, such a problem is of independent interest and, as far as we know, has not been previously considered.
In this paper, we consider the passage of a ion-acoustic solitary wave from an electron-ion plasma to a negative ion plasma and the reverse passage from a negative ion plasma  to an electron-ion plasma.

\section{FORMULATION OF THE PROBLEM}

We consider ion-acoustic solitary waves propagating in a collisionless plasma. For brevity, we will refer to them as solitary waves,  omitting the word ''ion-acoustic''.
This will not lead to a misunderstanding, since other types of waves are not considered here.
To describe the passage of solitary waves through a boundary between an electron-ion (EI) plasma and a negative ion (NI) plasma, we consider the motion of each of the three types of ions in both plasmas.
As for electrons, we do not distinguish them by belonging to one or another plasma.
We assume that all electrons are in equilibrium with the electric field, since in the problem under consideration the characteristic times are determined by the motion of ions.
In this case, the electron density can be determined by the Boltzmann formula.

In the EI plasma, we denote the unperturbed ion density and the unperturbed electron density as $ N_ {i0} $ and $ N_ {e0} $, respectively.
For the unperturbed density of positive ions, the unperturbed density of negative ions, and the unperturbed electron density in the NI plasma, we use the notation $ n_ {i0} $, $ n_ {j0} $ and $ n_ {e0} $, respectively.
It is obvious that in the unperturbed regions the conditions of quasineutrality are satisfied
    \begin{equation}
       Z_i N_{i0}=N_{e0},\qquad   z_i n_{i0}+z_j n_{j0}=n_{e0},\\
 \end{equation}
where $ Z_i> 0 $ is the charge number of ions in the EI plasma, and $ z_i> 0 $ and $ z_j <0 $ are the charge numbers of positive and negative ions in the NI plasma. 
We denote the distribution function of positive ions in the EI plasma  by $ F_ {i} (x, v, t) $, where $ x, v $ and $ t $ are the space, velocity and time variables, respectively.
The distribution functions of positive and negative ions in the NI plasma are denoted by $ f_ {i} (x, v, t) $ and $ f_ {j} (x, v, t) $, respectively. 
We are considering collisionless plasma.
The plasma motion is described by the Vlasov system of equations:
  \begin{equation}
   \begin{aligned}
    &\frac{\partial F_i}{\partial t}+v\frac{\partial F_i}{\partial x}-\mu_i\frac{\partial \varphi}{\partial x}\frac{\partial F_i}{\partial v}=0,\\
   &\frac{\partial f_i}{\partial t}+v\frac{\partial f_i}{\partial x}-\frac{\partial \varphi}{\partial x}\frac{\partial f _i}{\partial v}=0,\\
   &\frac{\partial f_j}{\partial t}+v\frac{\partial f_j}{\partial x}-\mu_j\frac{\partial \varphi}{\partial x}\frac{\partial f_j}{\partial v}=0,\\
    &\frac{\partial^2\varphi}{\partial x^2}=-(Z_iN_i +z_in_i+z_jn_j-n_{e}),\\
     &N_i=\int\limits_{-\infty}^{\infty}\!F_i(x,v,t)\,dv,\quad n_i=\int\limits_{-\infty}^{\infty}\!f_i(x,v,t)\,dv,\quad\\ &n_j=\int\limits_{-\infty}^{\infty}\!f_j(x,v,t)\,dv,\quad 
   n_e=n_{e0}\exp\varphi,
     \end{aligned}\label{e02}
 \end{equation}
 where 
 \begin{equation}
      \mu_i=\frac{Z_i m_i}{z_i M_i},\qquad \mu_j=\frac{z_j m_i}{z_i m_j} \label{e03}
 \end{equation}
are the parameters, characterizing the plasma ion composition. 
Here $\varphi$ is the electrostatic potential, $ M_i $ is the mass of the ion in the EI plasma, and  $ m_i $ and $ m_j $ are the masses of the positive ion and the negative ion in the NI plasma, respectively.
All quantities and equations are given in dimensionless form.
The quantities
 \begin{equation}
  \begin{aligned}
    &m_i, \quad n_{i0}, \quad  \left(\frac{m_i}{4\pi z_ie^2n_{i0}}\right)^{1/2},\quad \left(\frac{T_{e0}}{4\pi e^2n_{i0}}\right)^{1/2},\\
    &  \left(\frac{z_iT_{e0}}{m_i}\right)^{1/2},\quad
     z_iT_{e0}, \quad
    \frac{T_{e0}}{e},\quad
    n_{i0}\left(\frac{m_i}{z_i T_{e0}}\right)^{1/2}\\
    \end{aligned}\label{e04}
 \end{equation}
are used as units of mass, density, time, length, speed, temperature, potential and distribution function, respectively. 
Here $ e $ is the absolute value of the electron charge,  and  $ T_ {e0} $ is the constant  electron  temperature (in energy units).
As for ion temperatures, we restrict ourselves here to the case of cold ions, assuming that all ion temperatures are zero.

The initial conditions for the system of equations \eqref {e02} should take into account how the EI plasma  and the NI plasma are located relative to each other.
We assume that at the initial time $ t = 0 $ the boundary between the EI plasma  and the NI plasma is located at the point $ x = 0 $.
A solitary wave is created in the plasma, which is located in the region $ x \le 0 $ (region 1).
The initial conditions can be formulated for two cases.
The first case corresponds to the fact that region 1 is occupied by the EI plasma, and the solitary wave is formed in this plasma.
The region $ x \ge 0 $ (region 2) is occupied by the NI plasma.
In the second case, region 1 is occupied by the NI plasma, in which the solitary wave is formed, and region 2 is occupied by the EI plasma.
In both cases, the solitary wave in region 1 is created against the background of a uniformly distributed plasma with given unperturbed particle densities.
The plasma in region 2 at $ t = 0 $ has a uniform distribution with its given unperturbed particle densities.
Thus, at the initial  time, the density of each of the three types of ions falls from the unperturbed value to zero at the point $ x = 0 $, and the electron density here changes sharply from $ N_ {e0} $ in the EI plasma  to $ n_ {e0} $ in the NI plasma.
The solitary wave propagates in a positive direction.
The amplitude of the potential $ \varphi_ {m} $ varies from the initial amplitude $ \varphi_ {m1} $ in region 1 to  $ \varphi_ {m2} $ in region 2.
By $ \varphi_ {m2} $ we mean the stationary amplitude of a solitary wave in region 2 if such a solitary wave arises, or simply the amplitude of a disturbance if a solitary wave does not arise.

The expression for the electron density in Eq. \eqref {e02} is written in the form from which it follows that the potential $ \varphi (x, t) $ should be set equal to zero in the unperturbed region of the NI plasma, where $ n_e =n_ {e0} $.
The same formula shows that in the unperturbed region of the EI plasma, the potential is equal to $ \ln (N_ {e0} / n_ {e0}) $.
Therefore, at the initial  time, when both plasmas are unperturbed near the boundary, along with discontinuities in the densities of ions and electrons, at the point $ x = 0 $ there is a potential discontinuity determined by the discontinuity of electron density. 
  
Obviously, the presence of a discontinuity in the electron density and a discontinuity of the potential at the point $ x = 0 $  leads to their subsequent decays.
As a result, a disturbance  arises near the boundary between the two plasmas, in which a collisionless shock wave, a rarefactive wave, oscillations, and even instability can develop \cite {FP17}.
It is clear that in this case, the passage of the solitary wave through the boundary between the plasmas is substantially determined by the phenomena caused by the decay of the initial discontinuity.
In order to investigate the passage of a solitary wave through the boundary between the plasmas per se, it is necessary to choose the parameters of both plasmas so that the above-mentioned phenomena do not occur at this boundary.
It is easy to see that for this it is necessary to eliminate the discontinuity  in electron density by accepting
\begin{equation}
    N_{e0}=n_{e0}.
    \label{e05}
\end{equation}
Then the potential in the unperturbed region of the EI plasma  is $ \ln (N_ {e0} / n_ {e0}) = 0 $ and there is no potential discontinuity.
Accordingly, there are also no disturbances near the boundary between two plasmas.
Our numerical experiment confirms this \cite {FP17}.

In this article, the passage of a solitary wave through the boundary between two plasmas is studied numerically.
We consider the case when the positive ions in the EI plasma  and in the NI plasma are the same, and all the ions are singly charged, that is, $ m_i = M_i, \; z_i = Z_i = 1, \; z_j = -1 $.
The ratio of the mass of the negative ion to the mass of the positive ion is chosen to be $ m_j / m_i = 0.476 $, which takes place in a plasma with positive ions $ \mathrm {Ar ^ +} $ and negative ions $ \mathrm {F ^ -} $.
Parameters \eqref {e03} take the values $ \mu_i = 1, \; \mu_j = -2.101 $.
The dimensionless unperturbed ion densities in the NI plasma are chosen as follows: $ n_ {i0} = 1, \; n_ {j0} = 0.1 $.
To satisfy relation \eqref {e05}, we choose $N_ {i0} = N_ {e0} = n_ {e0} =0.9 $. 

Consider the ranges of possible amplitudes of solitary waves that can propagate in an  EI plasma  and in a NI plasma \cite{Sagdeev,Nakamura85c}.
In an EI plasma  with cold ions, solitary waves can have a potential amplitude  $ \varphi_ {m} $ in the range from 0 to the critical value $ \varphi_ {cr} = 1.256 $.
In a NI plasma, in addition to compressive solitary waves, rarefactive solitary waves can exist.
With our ratio of ion densities in the NI plasma,  the region of possible amplitudes of compressive solitary waves ($ \varphi_ {m}> 0 $) almost continuously passes into the region of possible amplitudes of rarefactive solitary waves ($ \varphi_ {m} <0 $).
An exception is a very small range of amplitudes near zero.
The amplitude of the compressive solitary wave cannot exceed the maximum  $ \varphi_ {mx} = 1.362$.
The negative amplitude of the rarefactive solitary wave cannot fall below the minimum  $ \varphi_ {mn} = - $ 0.512.

Since the ions are cold, the speed of sound in an EI plasma  is $ C_0 = 1 $, and in a NI plasma the speed of sound for the only possible mode of ion-acoustic oscillations is determined by the formula
\begin{equation}
c_0=\left(\frac{z_in_{i0}+z_j\mu_j n_{j0}}{z_in_{i0}+z_jn_{j0}}\right)^{1/2}.
    \label{e06}
 \end{equation} 
In the case of $ n_ {j0} = 0.1 $, we have $ c_0\approx 1.1596 $. 
For a given ionic composition, the propagation velocity of a solitary wave in an EI plasma  $ D $ and the propagation velocity of a solitary wave in a NI plasma $ d $ depend only on the solitary wave amplitude if the ions are cold.
The dependencies $ D (\varphi_ {m}) $ and $ d (\varphi_ {m}) $ are shown in Fig. 1.
It also shows the speed of sound  in the NI plasma.
It is seen that the propagation velocity of a solitary wave with an amplitude of $ \varphi_ {m} <0.429 $ in an EI plasma is lower than the speed of sound in the NI plasma considered here.
It can be expected that this has a certain effect on the passage of such a solitary wave from the EI plasma  to the NI plasma.
In addition, as can be seen from Fig. 1, if, after the passage from one plasma to another, the compressive  solitary wave is retained as a compressive solitary wave with a slightly changed propagation velocity $ d \sim D $, then the amplitude of the solitary wave in the EI plasma  is always greater than its amplitude in the NI plasma.
But such a situation cannot occur during the passage of a solitary wave of large amplitude $ (\varphi_ {m} \gtrsim 1.256) $ from the NI plasma to the   EI plasma, since the amplitude of the solitary wave in the EI plasma  cannot increase above the critical value $ \varphi_ {cr} = 1.256 $.
In other words, the case of a solitary wave of large amplitude requires a separate consideration.
A separate consideration is also necessary for the case when a rarefactive solitary wave created in the NI plasma enters the EI plasma, where a rarefactive solitary wave cannot propagate.
We are considering several options for formulation of the problem, which differ both in the initial amplitudes of the solitary wave and in the arrangement of the EI plasma  and the NI plasma relative to each other.

\begin{figure}\centering
	\includegraphics[viewport= 179 650 401 790, width=222pt]{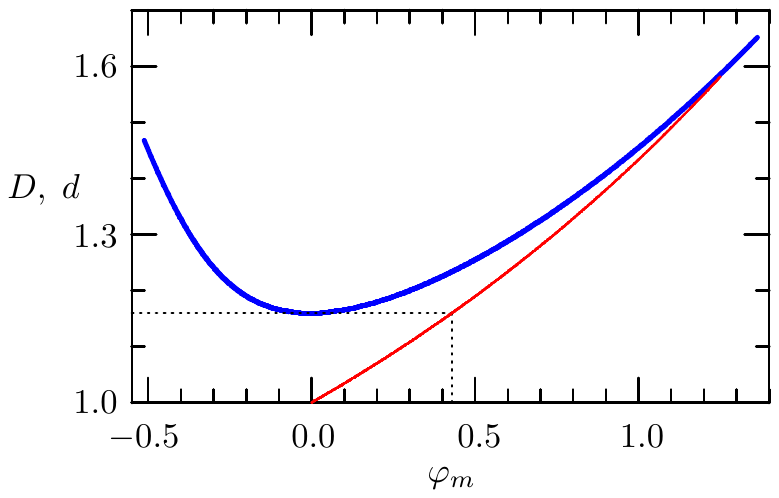}
	\caption{The propagation velocities of solitary waves in the EI plasma  $ D $ (solid red curve) and in the NI plasma $ d $ (thick blue curve) depending on the amplitude $ \varphi_m $. The horizontal dotted curve shows the speed of sound in the NI plasma.
} \label{Fig1}
\end{figure}

The problem is solved by numerical simulation using the particle-in-cell (PIC) method.
Numerical simulation corresponds to the system of equations \eqref {e02}.
As a rule, the length of region 1 is chosen equal to 80, with the exception of cases of solitary waves with small amplitudes, when the length of region 1 should be increased.
We call the point $x = a$ the coordinate of the solitary wave if $\varphi (a) = \varphi_m$.
As a rule, the initial coordinate of the solitary wave is $ x = -40 $.
The length of region 2 varies up to several thousand, depending on the specific goals of the calculation.
On the left boundary of region 1 and on the right boundary of region 2, the electric field is set to zero.
At the same boundaries, the specular reflection condition is specified for particles.
The movement of ions of each sort is simulated by a separate group of particles.
Typically, the total number of particles is about $5 \times 10^7$. 
The number of particles for each species is set in accordance with the density and size of the region occupied by the simulated plasma component.

To correctly set the initial spatial distributions of quantities in a solitary wave, we find exact solutions for the distributions of the densities and velocities of particles at a given amplitude using the method described previously \cite{FP09,Book}.
These distributions are implemented in the PIC code using  an appropriate set of  coordinates and velocities of particles. 
In all our numerical experiments, a solitary wave of any amplitude formed in this way propagates without any changes in its amplitude, propagation velocity and shape.

\section{NUMERICAL SIMULATION RESULTS}

\subsection{Passage of a solitary wave from an electron-ion plasma into a negative ion plasma}

\begin{figure}[t]\centering
	\includegraphics[viewport= 194 374 415 792,width=221pt]{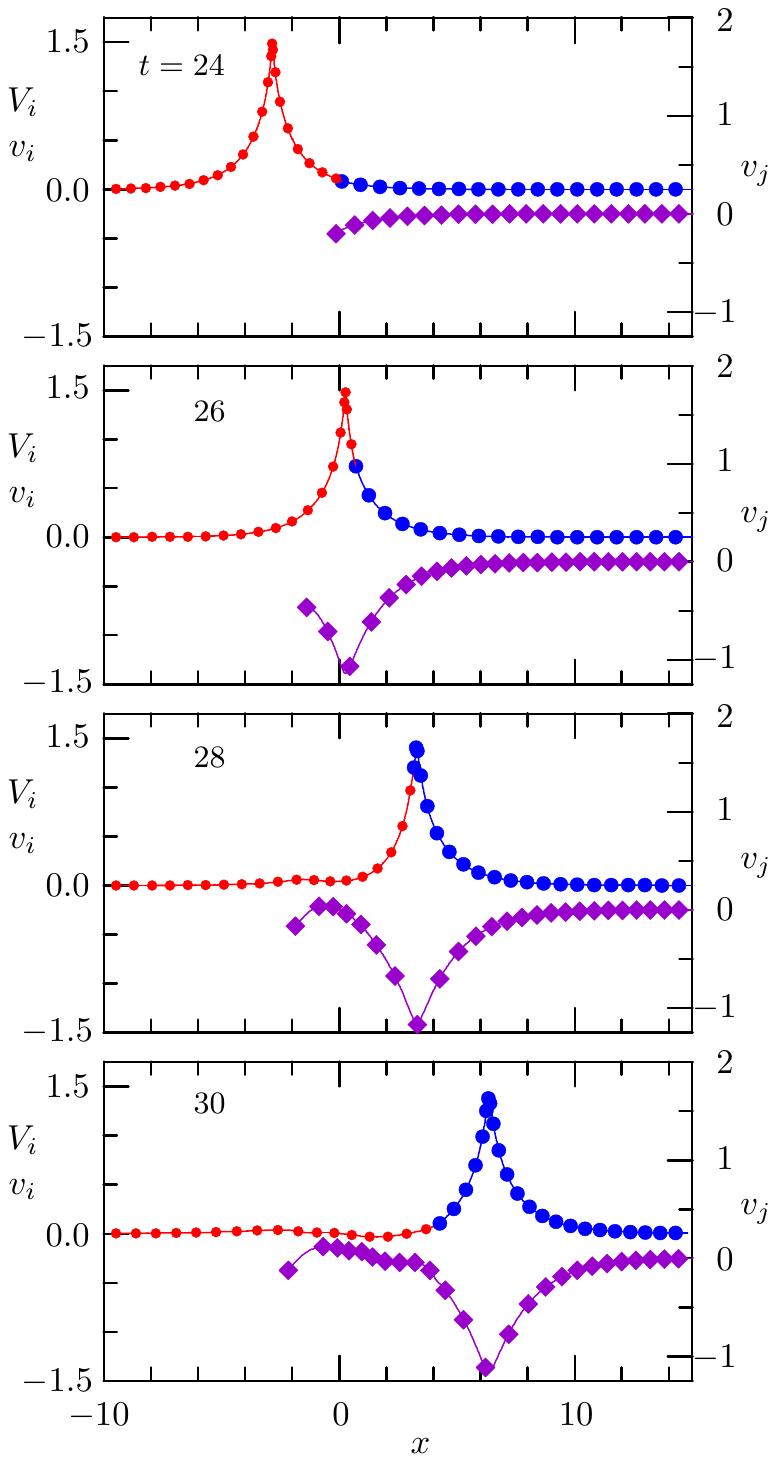}
	\caption{The phase planes of positive ions in the EI plasma  (red circles) and in the NI plasma (large blue circles), as well as the phase planes of the negative ions in the NI plasma (violet rhombs, right axis) at different times. $ \varphi_ {m1} = 1.2$.} \label{Fig2}
\end{figure}

Let us first consider the passage of a solitary wave from an electron-ion plasma into a negative ion plasma.
We illustrate the passage by the example of a solitary wave with the initial amplitude of the potential $ \varphi_ {m1} = 1.2 $ and the propagation velocity $ D = 1.550 $.
The initial coordinate of the solitary wave is $ x = -40 $.
Figure 2 shows the phase planes of positive ions $ (x, V_i) $ in the EI plasma, positive ions  $(x,  v_i) $ and negative ions  $(x, v_j) $ in the NI plasma for different times.
(Note that in this and the following figures, we use different colors to represent data related to different plasma components.
Data describing dependencies in the EI plasma  region are shown in red.
Data related to the NI plasma region, including positive ion data in this plasma, are shown in blue.
Data related to negative ions in the NI plasma are shown in violet.)

Figure 2 shows that, at $ t = 24 $, the solitary wave, which is still in the EI plasma, approaches the boundary with the NI plasma $ x = 0 $ and begins to enter  this plasma.  
Positive and negative ions of the NI plasma acquire velocities and shift under the influence of the solitary wave field.   
Subsequently, the solitary wave gradually enters the region of the NI plasma.    
The velocities of positive ions in the EI plasma after the passage of the solitary wave decrease to zero. 
 But at the same time, some positive ions from the EI plasma, close to the boundary, enter the region $ x> 0 $, originally occupied by  the NI plasma.   
Positive ions from  the NI plasma give way to them and themselves shift in the direction of propagation of the solitary wave.
Negative ions from  the NI plasma also shift, but in the opposite direction, so that some negative ions close to the boundary go into the region $ x <0 $.             
These ion shifts are an obvious consequence of the unipolarity of solitary waves.    

\begin{figure}\centering.
	\includegraphics[viewport= 198 598 415 792, width=217pt]{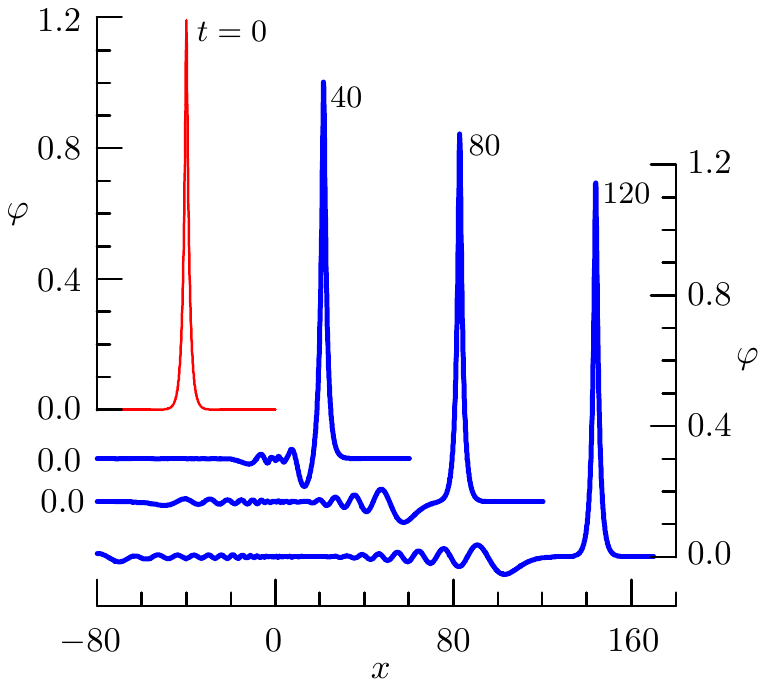}
	\caption{Potential distributions at different times in the case of $ \varphi_ {m1} = 1.2 $. To avoid overlapping, the vertical axes for the distributions at $ t = 40,  80$ and $ 120 $ are shifted down. Zero values on the vertical axes for distributions at $ t = 40 $ and $ t = 80 $ are shown on the left. The vertical axis for the distribution at $ t = 120 $ is shown on the right.} \label{Fig3}
\end{figure}

The evolution of the distribution of potential over time can be seen in Fig. 3.
At $ t = 40 $, the solitary wave in the NI plasma is still being formed. 
A disturbance consisting of oscillations forms behind the front peak, and the first oscillation, which has the largest amplitude, is negative.
The disturbance propagates at a subsonic speed and by the time $ t = 120 $  is completely separated from the front peak, which now represents a steady solitary wave.
The amplitude of the solitary wave potential is  $ \varphi_ {m2} \approx 1.146 $, and the solitary wave propagates at the velocity of $ d \approx 1.529 $.
As we see, the solitary wave passing from the EI plasma to the NI plasma  remains a solitary wave, but with a slightly lower amplitude, as was suggested above.

As mentioned in the previous section, the propagation velocity of a solitary wave with an initial amplitude $ \varphi_ {m1} <0.429 $ in the EI plasma is less than the speed of sound in the NI plasma.
To consider the passage of a solitary wave with such an amplitude from the EI plasma to the NI plasma, we performed a numerical simulation of the passage for the case $ \varphi_ {m1} = 0.4 $.
The calculation shows that the solitary wave excites in the NI plasma a soliton-like disturbance  of a noticeably smaller amplitude $ \varphi_ {m2} \approx 0.310 $, propagating with a not very high supersonic velocity $ d \approx 1.202 $.
The disturbance is followed by oscillations propagating at a speed less than the speed of sound  $ c_0 \approx1.1596 $.
Since the difference in these velocities is not very large, a sufficiently large time is required for the separation of the front disturbance from the oscillations.

\begin{figure}\centering
	\includegraphics[viewport= 191 652 402 790,width=211pt]{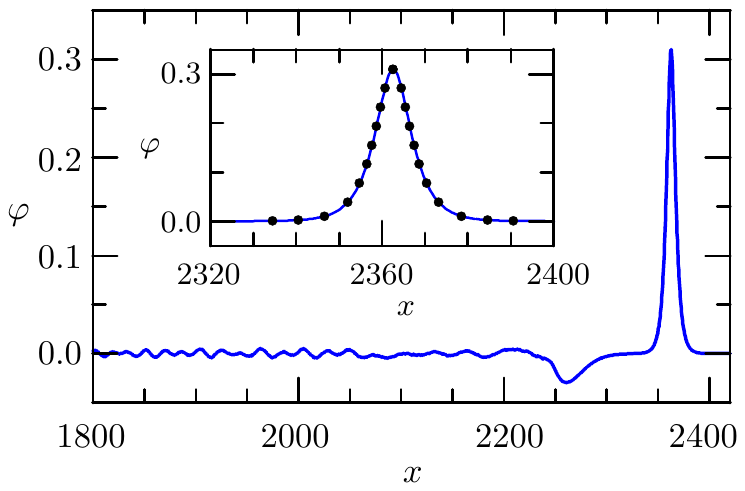}
\caption{Potential distribution in the NI plasma at $ t = 2000 $ in the case of $ \varphi_ {m1} = 0.4 $.  In the inset, the distribution obtained in the numerical experiment (solid blue curve) is compared with the theoretical profile of a solitary wave (circles) with the same amplitude $\varphi_{m2}\approx 0.310$.}
\label{Fig4}
\end{figure}

Figure 4 shows the potential distribution in the NI plasma at $ t = 2000 $, when such a complete separation occurred.
To make sure that the formed soliton-like disturbance is exactly a solitary wave corresponding to the obtained amplitude $ \varphi_ {m2} \approx  0.310$, we calculated the potential profile of a solitary wave with the same amplitude using the method described in Refs.  \onlinecite{FP09,Book} 
and compared it with the profile obtained in the numerical experiment.
The potential distributions obtained by two methods are shown in the inset in Fig. 4.
As can be seen, there is a complete coincidence of the two profiles.
Thus, even with a relatively small amplitude of the solitary wave $ \varphi_ {m1} $ in the EI plasma, its passage into the NI plasma leads to the formation of a solitary wave there. 

\begin{figure}\centering
	\includegraphics[viewport= 192 650 412 793,width=220pt]{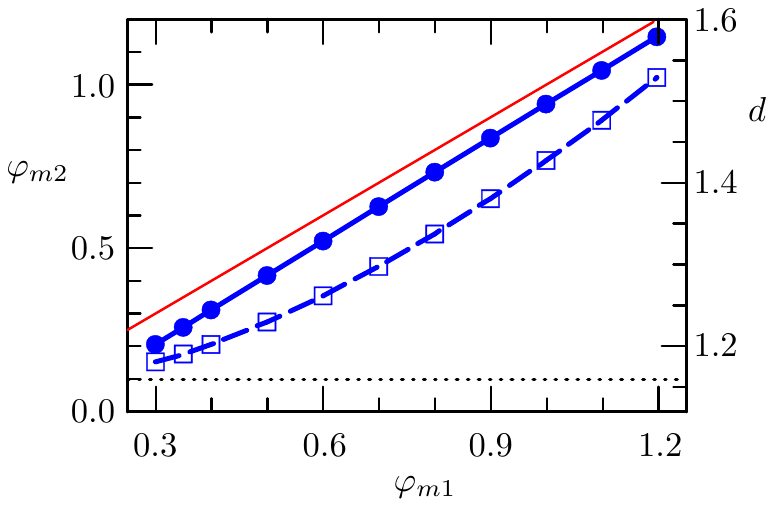}
	\caption{The amplitude of the potential $ \varphi_ {m2} $ (blue solid curve and circles) and the propagation velocity $ d $ (blue dashed curve and squares, right axis) of the solitary wave in the NI plasma depending on the initial amplitude of the potential $ \varphi_ { m1} $ in the EI plasma. Symbols show the results of numerical experiments, the thin red line is $ \varphi_ {m2} = \varphi_ {m1} $. The dotted curve shows the speed of sound in the NI plasma (right axis).} 	\label{Fig5}
\end{figure}
 
Fig. 5 summarizes the results of numerical experiments. 
It depicts the amplitude $ \varphi_ {m2} $, as well as the propagation velocity $ d $ of a stable solitary wave in the NI plasma, depending on the initial amplitude of this solitary wave in the EI plasma $ \varphi_ {m1} $.
To estimate the deviation of the amplitude in the NI plasma  from the initial amplitude in the EI plasma, the line $ \varphi_ {m2} = \varphi_ {m1} $ is plotted in Fig. 5.
It can be seen that the amplitude of a solitary wave in the NI plasma is always less than the initial amplitude in the EI plasma.

\subsection{Passage of a compressive solitary wave from a~negative ion plasma into an~electron-ion plasma}

In a NI plasma, a solitary wave can be  a compressive wave $ (\varphi_m> 0) $ or a rarefactive wave $ (\varphi_m <0) $.
In this subsection, we consider  the passage of a compressive solitary wave formed in a NI plasma  into an EI plasma.
As before, the boundary between the plasmas is located at the point $ x = 0 $.
The characteristic features of the passage of a solitary wave through the boundary between two plasmas can be seen in Fig. 6, where, as an example, the case $\varphi_{m1}=0.8$  is presented.
Here, both the distributions of the potential at different times and the dependence of the amplitude of the solitary wave on its coordinate are given.
We see that at $ t = 20 $ the solitary wave propagating in the NI plasma approaches the boundary between two plasmas $ x = 0 $.
Subsequently, its amplitude in a short time first increases, then rapidly drops, and finally begins to increase slowly up to a stationary value in the EI plasma.
The potential distribution is characterized by the presence of a front peak, followed by a disturbance consisting of oscillations.
We note that here the first oscillation, which has the largest amplitude, is positive, in contrast to the negative first oscillation in the NI plasma (see Fig. 3).
It is clearly seen that with time the amplitude grows above the initial value in the NI plasma and at $ t = 110 $ reaches the stationary value $ \varphi_ {m2} \approx 0.848$.
The front peak is completely separated from the oscillations and propagates with a constant amplitude and velocity, that is, it is a solitary wave in the EI plasma.
\begin{figure}[t]\centering
	\includegraphics[viewport= 186 652 399 790,width=213pt]{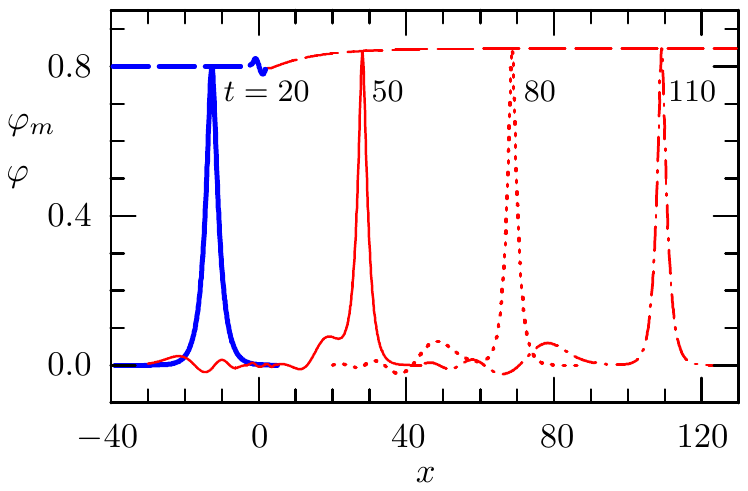}
	\caption{Potential distributions at different times and the dependence of the potential  amplitude on the coordinate $ \varphi_m (x) $ (dashed line) at $ \varphi_ {m1} = 0.8 $.}  \label{Fig6}	
\end{figure}
  
Above, we drew attention to the case of a solitary wave with a sufficiently large initial amplitude $ \varphi_ {m1} $. Upon the passage of such a solitary wave from the NI plasma into the EI plasma, the  increase in the amplitude in the EI plasma may not occur, since here the amplitude is limited by the critical value $ \varphi_ {cr} =  1.256$.
To study this situation in detail, we performed 3 numerical experiments in which the amplitude $ \varphi_ {m1} $ was large. The results of these experiments are presented in Fig. 7.    

It can be seen that all of the presented quantities undergo fairly rapid changes when passing through the boundary between the plasmas, and then gradually reach constant values.
For an initial amplitude of $ \varphi_ {m1} = 1.2 $ in the NI plasma  (Fig. 7(a)), the amplitude of the solitary wave in the EI plasma does not exceed the critical value.
The amplitude of the ion velocity in the EI plasma $ V_m $ approaches the propagation velocity of the solitary wave $ D $, but does not exceed it.
\begin{figure}[h]\centering
	\includegraphics[viewport= 192 407 414 793,width=222pt]{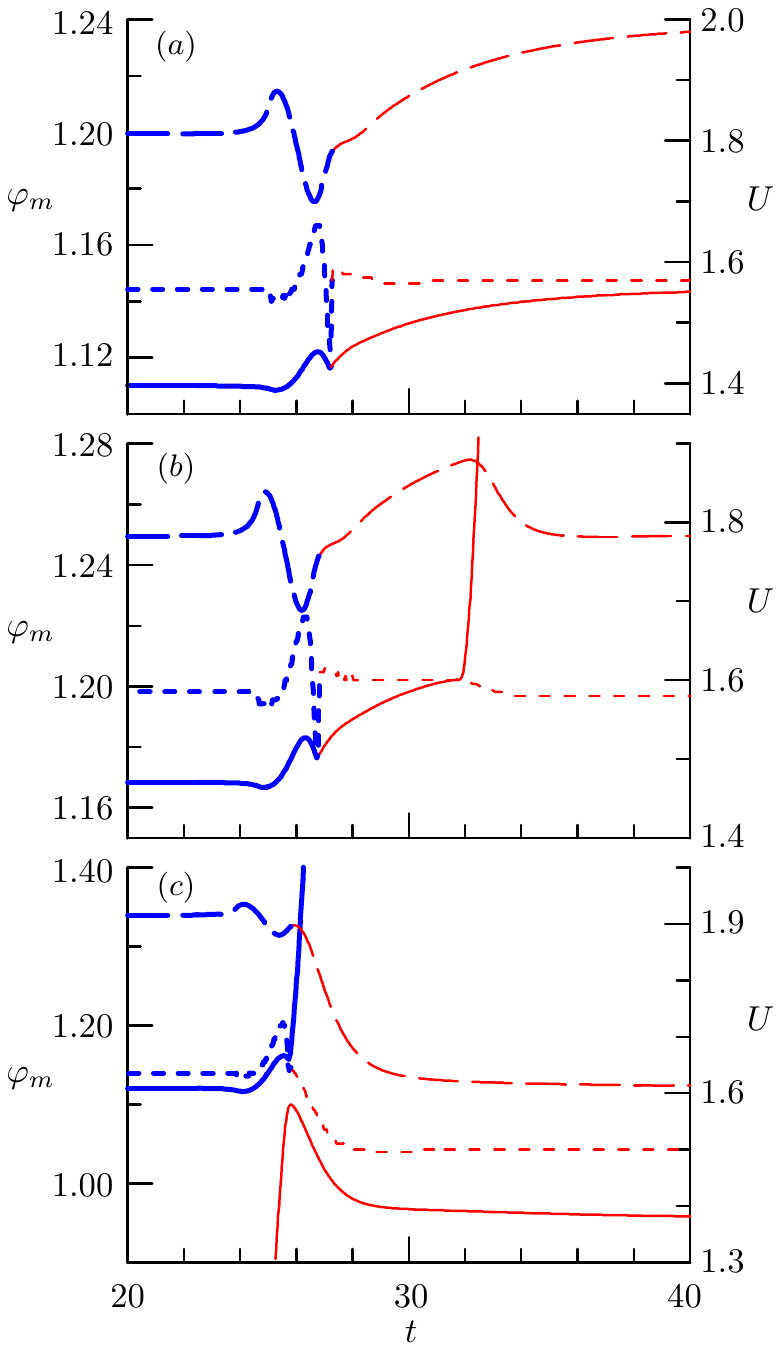}
	\caption{The amplitudes of the potential $ \varphi_m $ (long-dashed curves),
 the propagation velocities of the solitary wave $ d $ in the NI plasma and $ D $ in the EI plasma (short-dashed curves,  right axis), and also the amplitudes of the velocities of positive ions $ v_m $ in the NI plasma and $ V_m $ in the EI plasma (solid curves,  right axis) as  functions of time for various initial amplitudes: $ (a) $ $ \varphi_ {m1} = 1.2 $, $ (b) $ $ \varphi_ {m1} = 1.25 $, $ (c) $ $ \varphi_ {m1} = 1.34 $.  Here we denote $U=\{d, D, v_m, V_m\}$. The quantities related with the NI plasma are shown by thick blue curves, and the quantities related with the EI plasma are shown by red curves.	
} \label{Fig7}
\end{figure}

At a higher initial amplitude $ \varphi_ {m1} = 1.25 $ (Fig. 7(b)), the amplitude of the  solitary wave in the EI plasma increases with time above the critical value $ \varphi_ {cr} = 1.256 $.
Simultaneously with an increase in the amplitude of the potential, the amplitude of the ion velocity in the EI plasma $ V_m $ also increases, and at some point in time this amplitude becomes greater than the propagation velocity of the solitary wave $ D $.
This leads to the separation of all ions that have gained velocity in the solitary wave field into two groups.
The main group consists of ions whose velocities are less than $ D $.
 These ions are in the region of the decelerating electric field and stop over time.
Another group consists of ions whose velocities are greater than $ D $.
These ions overtake the peak of the solitary wave potential and enter the region of an accelerating electric field.
In this field, the ions very quickly acquire even greater velocities (the maximum velocity, not shown in Fig. 7(b), $ V_m = 3.441 $), leave the solitary wave region and then move along the unperturbed plasma.
Note that in this case the initial amplitude of the velocity of positive ions in the NI plasma $ v_m = 1.471 $ is less than the maximum permissible propagation velocity of the solitary wave in the EI plasma $ D_ {cr} = 1.585 $. 

A somewhat different situation arises in the case of an even larger initial amplitude $ \varphi_ {m1} = 1.34 $ (Fig. 7(c)).
Here, both the initial amplitude of the velocity of positive ions, equal to $ v_m = 1.611 $, and  the initial velocity of propagation of the solitary wave, $ d = 1.638 $, are greater than $ D_ {cr} $ in the EI plasma.
Therefore, when the solitary wave enters the EI plasma, its propagation velocity should immediately begin to fall, and now other ions,  positive ions from the NI plasma, overtake the solitary wave peak.
 In the accelerating electric field of the solitary wave, these ions very quickly acquire even greater velocities (the maximum velocity, not shown in Fig. 7(c), $ V_m = 3.873 $) and then continue to move along the unperturbed region.  

In fact, the observed ion acceleration is a breaking of the ion velocity profile.
Figure 8 shows how this process develops over time.
Here we see the separation of the group of fast ions from the bulk of the ions and their subsequent acceleration.
Naturally, the acceleration of ions occurs due to the energy of the solitary wave.
Using this mechanism, the solitary wave throw off excess energy, and its amplitude is set to an acceptable level.

\begin{figure}\centering
	\includegraphics[viewport= 189 596 399 790,width=210pt]{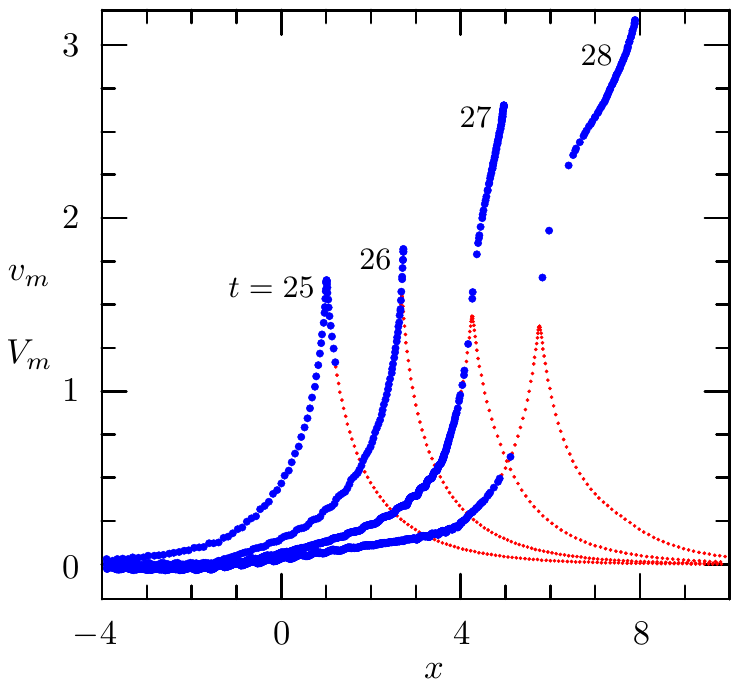}
	\caption{The phase planes of positive ions in the NI plasma  (large blue circles) and in the EI plasma (small red circles) at different times for the case $ \varphi_ {m1} = 1.34 $.} \label{Fig8}	
\end{figure}

\begin{figure}\centering
	\includegraphics[viewport= 192 650 413 790,width=221pt]{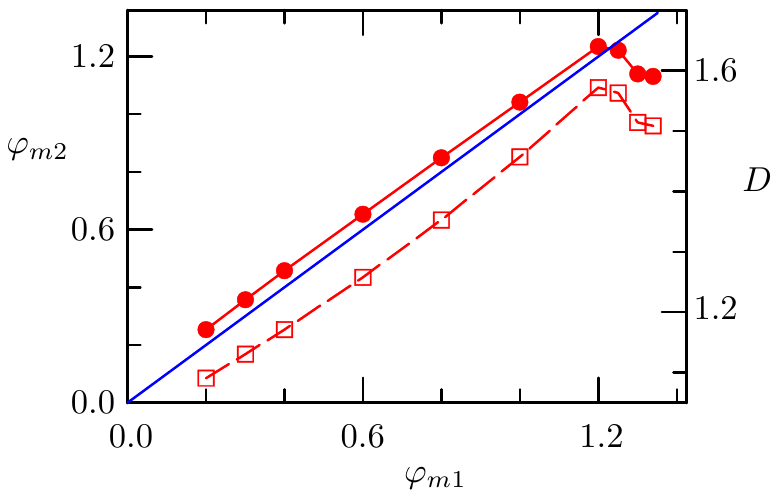}
	\caption{The amplitude of the potential $ \varphi_ {m2} $ (red solid curve and circles) and the propagation velocity $ D $ (red dashed curve and squares, right axis) of the solitary wave in the EI plasma depending on the initial amplitude of the potential $ \varphi_ { m1} $ in the NI plasma. Symbols show the results of numerical experiments, the thin blue line is $ \varphi_ {m2} = \varphi_ {m1} $.} \label{Fig9}	
\end{figure}

Figure 9 shows the dependences of the amplitude $ \varphi_ {m2} $ and the propagation velocity $ D $ of a stable solitary wave in the EI plasma on the initial amplitude of this solitary wave $ \varphi_ {m1} $ in the NI plasma.
Plotted on Fig. 9, the line $ \varphi_ {m2} = \varphi_ {m1} $ allows us to estimate the deviation of the amplitude of a solitary wave in the EI plasma from the initial amplitude in the NI plasma.
It is seen that in the amplitude range $0<\varphi_{m1}\le 1.2$, the amplitude of a solitary wave in the EI plasma is higher than the corresponding initial amplitude in the NI plasma.
However, in the case of a solitary wave with a higher initial amplitude in the NI plasma, its stable amplitude in the EI plasma is less than the initial amplitude due to the above-described mechanism for reducing the energy of a solitary wave.

\subsection{Passage of a compressive solitary wave through a layer of negative ion plasma}

Above, we discussed changes in the amplitude of the potential of a solitary wave during its passage through the boundary between two plasmas.
Examples of the dependence of the amplitude on time are shown in Fig. 7, and an example of the dependence of the amplitude on the coordinate of a solitary wave, which also varies with time, is given in Fig. 6.
It is of interest to compare the changes in the amplitude of a solitary wave that occur during its passage from an EI plasma to a NI plasma, with those changes that occur during a reverse passage from a NI plasma to an EI plasma.
This is convenient to do in numerical experiments simulating the case when both passages are present simultaneously.
We performed such numerical experiments by simulating the case when there are three plasma layers: a layer of EI plasma, a layer of  NI plasma, and again a layer of EI plasma.

The problem is formulated as follows.
At the initial  time, in the region $ -80 \le x \le 0 $ there is an EI plasma in which there is a solitary wave located at the point $ x = -40 $.
The solitary wave propagates in the direction of the region $ 0 \le x \le 240 $, where a  NI plasma is located.
After passing through the region of NI plasma, the solitary wave enters the region of $ 240 \le x \le 560 $, where another EI plasma is located.
All parameters of the NI plasma and the parameters of both EI plasmas are exactly the same as the corresponding parameters of the NI plasma and the EI plasma in previous calculations.

The change in the amplitude of a solitary wave $ \varphi_ {m} $ over time as it passes from the EI plasma to the NI plasma and again to the EI plasma at different initial amplitudes $ \varphi_ {m1} $ is shown in Fig. 10.
Here we consider rather large initial amplitudes $ \varphi_ {m1} $, since in such cases a solitary wave in the NI plasma takes its final shape before it passes through the entire plasma region  240 in length.
From Fig. 10, it can be seen that in the first and second passages, the amplitude of a solitary wave changes rapidly.
Such a change occurs mainly in the region located to the left of the boundary between the plasmas.
The spread of $ \varphi_ {m} $ in the regions of sharp changes is approximately the same for both passages.
It is seen that the amplitude of a solitary wave near the boundary between the plasmas first decreases, and then increases from the side of the EI plasma.
 From the side of the NI plasma, the amplitude of a solitary wave first increases and then decreases.
 
\begin{figure}[t]\centering
	\includegraphics[viewport= 185 531 400 790,width=215pt]{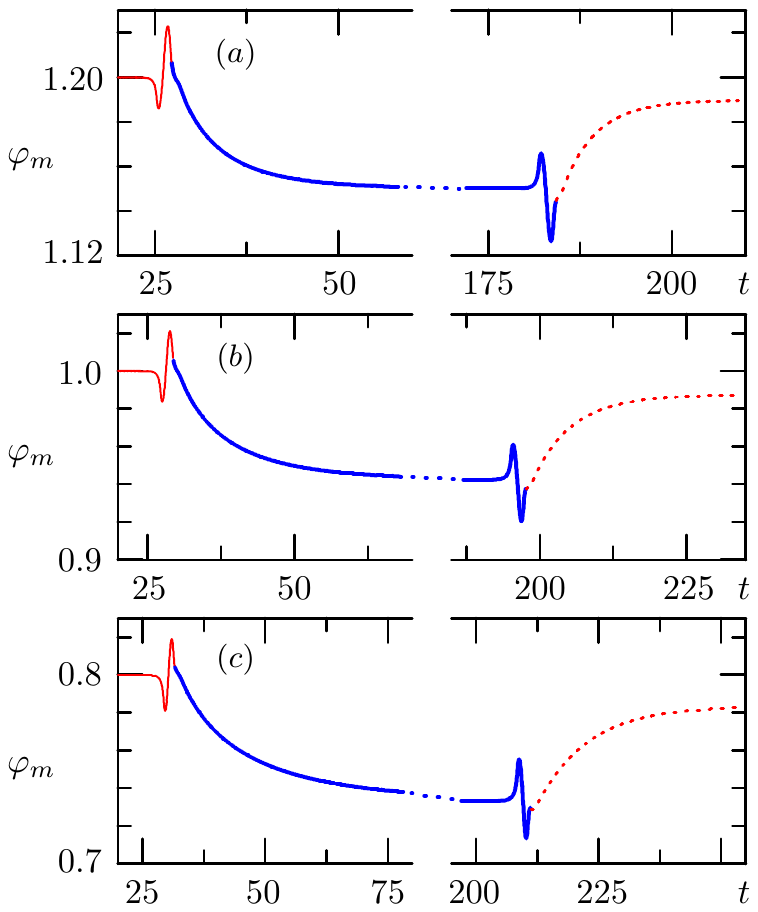}
	\caption{The potential amplitudes $ \varphi_ {m} $ as  functions of time  in the first region of the EI plasma (thin red curves), in the region of NI plasma (thick blue curves) and in the second region of the EI plasma (red dotted curves) for different $ \varphi_ {m1} $: (a) $ \varphi_ {m1} = 1.2 $, (b) $ \varphi_ {m1} = 1.0 $, (c) $ \varphi_ {m1} = 0.8 $. 
} \label{Fig10}	
\end{figure}

\begin{figure}[h]\centering
	\includegraphics[viewport= 192 624 404 790,width=212pt]{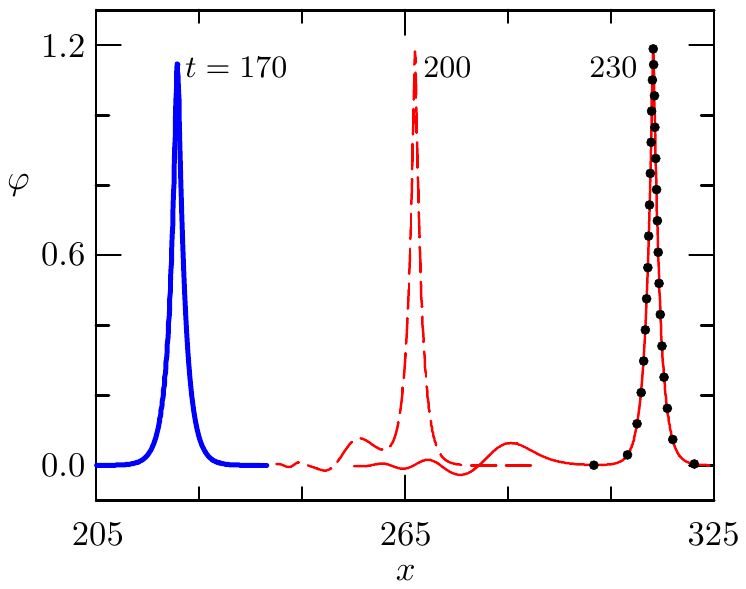}
	\caption{Potential distributions in the NI plasma region at $ t = 170 $ (thick blue curve) and in the second region of  the EI plasma at $ t = 200 $ and $ t = 230 $ (red curves) in the case of  $ \varphi_{m1} = 1.2 $. The circles show the theoretical profile of a solitary wave in an EI plasma at $ \varphi_m = 1.190 $ and $ D = 1.544 $.} \label{Fig11}	
\end{figure}  

From Fig. 10 it is clearly seen that the amplitude of the solitary wave in the second region of the EI plasma is less than its initial amplitude in the first region of the EI plasma.
The decrease is due to the fact that a part of the energy of a solitary wave during passages is spent on exciting oscillations following the main front peak, from which a solitary wave is formed over time.
This occurs after the separation of the oscillations propagating at a subsonic speed from the main peak.
Fig. 11 illustrates this process.
It can be seen that by the time  $ t = 230 $, a solitary wave with a stable amplitude of $ \varphi_ {m2} \approx 1.190 $ is formed in the second EI plasma.
The profile of a solitary wave is completely determined by its amplitude.
Knowing the amplitude obtained in the numerical experiment, we can calculate the theoretical profile of the potential \cite {FP09,Book} that should be in this solitary wave and compare it with the profile from the numerical experiment.
A comparison of two profiles is shown in Fig. 11, where their good coincidence is visible.
Thus, a solitary wave after passing through a layer of NI plasma remains a solitary wave with the correct potential profile, corresponding to a somewhat reduced amplitude.

\subsection{Passage of a rarefactive solitary wave from a negative ion plasma  into an electron-ion  plasma}

\begin{figure}[t]\centering
	\includegraphics[viewport= 194 411 409 794,width=215pt]{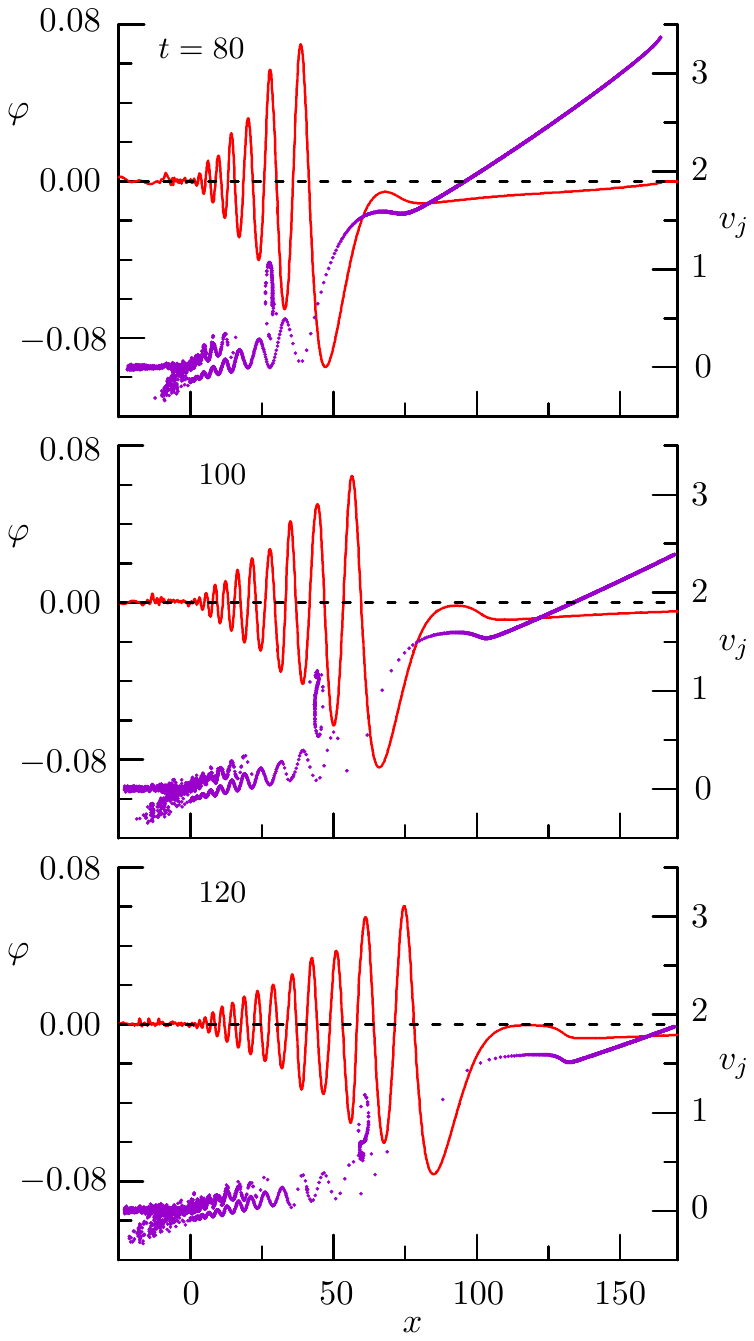}
	\caption{Potential distributions (red curves) and phase planes of negative ions (violet symbols, right axis) in the EI plasma at different  times. The dashed line denotes $\varphi=0$.}
	 \label{Fig12}
\end{figure}

So far, we have considered the passage of only compressive solitary waves $ (\varphi_m> 0) $ through the boundary between two plasmas.
In a NI plasma, a rarefactive solitary wave $ (\varphi_m <0) $ can exist.
But in an EI plasma, rarefactive solitary waves cannot propagate.
Therefore, a rarefactive solitary wave propagating through a NI plasma  does not remain a solitary wave when passing through a boundary with an EI plasma.
Obviously, some disturbance arises in the EI plasma, the evolution of which we consider below.
We illustrate the results obtained by an example of the passage of a rarefactive solitary wave with an amplitude $ \varphi_ {m1} = - 0.5 $ through the boundary between the plasmas.

The general idea of the processes occurring in an EI plasma is given in Fig. 12. 
We can see that an disturbance consisting of oscillations arises in the EI plasma.
The disturbance propagates in a positive direction at a subsonic speed.
It begins with a negative half-wave with a maximum amplitude, followed by a series of oscillations.
The amplitude of each  following oscillation decreases compared with the amplitude of the foregoing one in such a way that near the boundary with the NI plasma  $ x = 0 $ the oscillations almost completely disappear.
As the oscillations propagate, their number increases, and the amplitudes decrease.

An analysis of the results of numerical simulation shows that the most important features of the process under consideration are associated with the movement of negative ions of the NI plasma.
When a rarefactive solitary wave propagates, negative ions shift in the direction of propagation of the solitary wave.
Therefore, when a solitary wave approaches the boundary with an EI plasma, some negative ions enter the region occupied by this plasma.
These are the ions that were originally located near the boundary between the plasmas.
From Fig. 12 it can be seen that some negative ions acquire sufficiently high velocities and move ahead of the oscillations.
In addition, there is a group of negative ions that are located in the region of the second positive half-wave of the oscillations and move with them, that is, they are trapped by these oscillations.
And finally, there are negative ions with small velocities that oscillate in the disturbance  field.

\begin{figure}[t]\centering
	\includegraphics[viewport= 206 411 418 793,width=212pt]{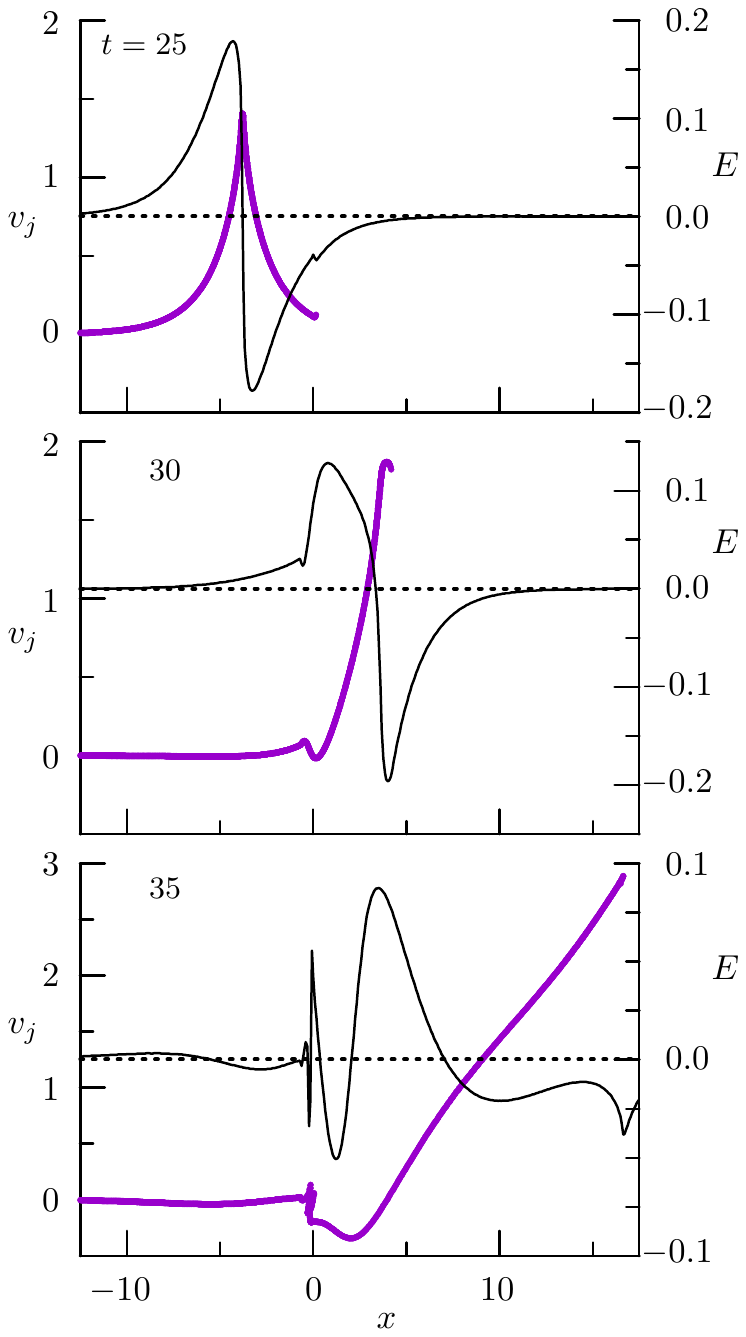}
	\caption{Phase planes of negative ions (violet symbols) and electric field distributions $E(x)$ (black curves, right axis) at different  times.  The dotted line denotes $E=0$.}
	 \label{Fig13}
\end{figure}

Let us first consider  the process of acceleration of negative ions. 
From Fig. 13(a) it can be seen that at $ t = 25 $ the solitary wave approaches the plasma boundary $ x = 0 $. 
The spatial distributions of the electric field $E$ and the velocities of negative ions practically do not differ from their initial distributions.
Subsequently, the solitary wave enters the region of the EI plasma, and near the boundary  disturbances arise both in the EI plasma and in the NI plasma.
The profile of the electric field is distorted.
At the same time, negative ions located near the boundary $ x = 0 $ and set in motion by the solitary wave also enter the region of the EI plasma.
Already at $ t = 30 $, such ions appear in the region $ x> 0 $ (Fig. 13(b)).
Some of these ions fall into the region of the negative electric field and are accelerated.
Accelerated negative ions continue to move in the region of the negative electric field, acquire even greater velocities and overtake the oscillations, forming a group of accelerated ions (Fig. 13(c)).

Fig. 12 shows that the disturbance arising in the EI plasma can trap a certain group of negative ions and transfer them in the region of the EI plasma over a rather long distance.
To describe in more detail the formation of trapped negative ions and their motion, we traced the motion of ten particles simulating negative ions.
As might be expected, there are three characteristic particle trajectories.
Fig. 14 presents data for three particles with different initial coordinates.
Shown here are the time dependences of the coordinates and velocities of the particles, as well as the electric field at the points where the particles are.

\begin{figure}\centering
	\includegraphics[viewport= 195 409 412 790,width=221pt]{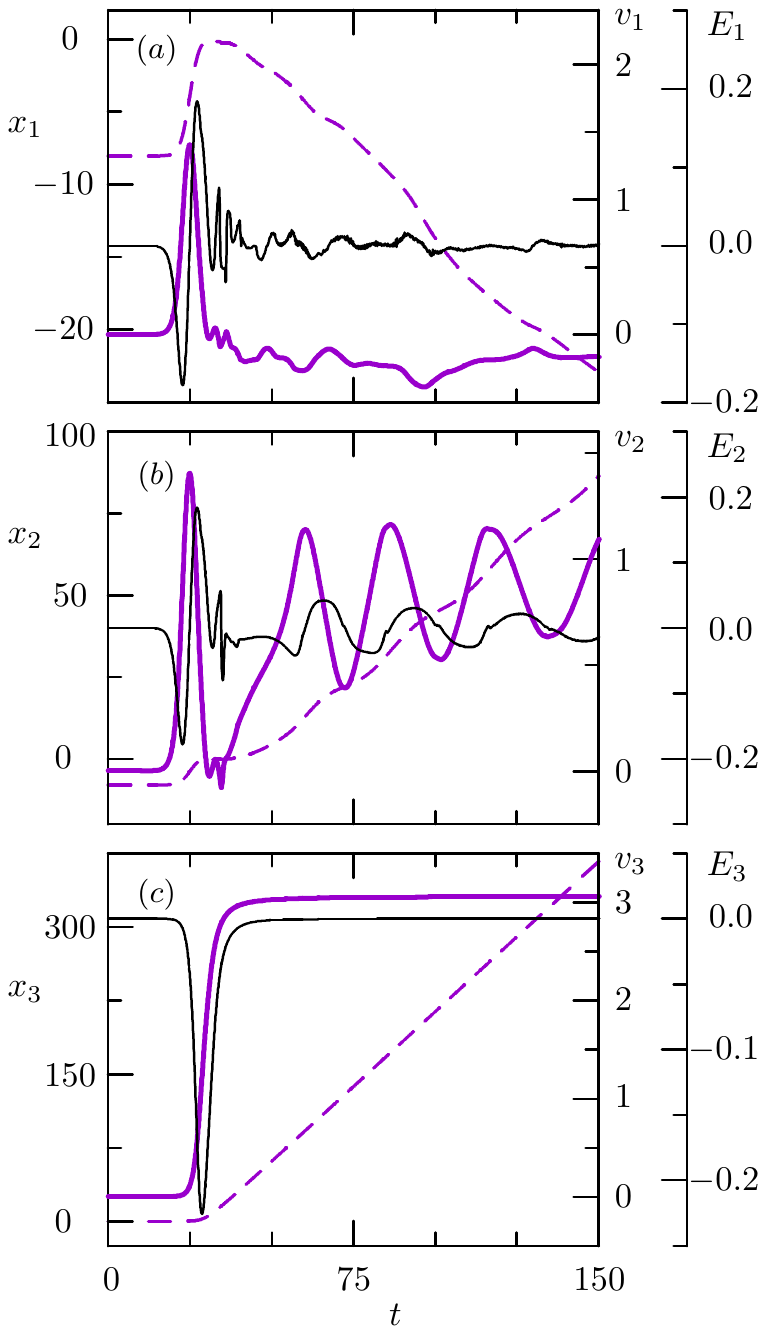}
	\caption{Coordinates $x_k$ (violet dashed curves), velocities $v_k$ (thick violet curves, right axis) and electric fields $E_k$ (black curves, separate right axis) as functions of time for three particles $(k=1, 2, 3)$.  The particles have the initial coordinates: $ x_ {1} = - 8.0220$,   $x_ {2} = - 8.0396$  and $x_ {3} = - 0.8816$.}
	 \label{Fig14}
\end{figure}

From Fig. 14(a), it can be seen that  particle 1 does not fall into the region $ x> 0 $ and gradually moves to the inner part of the NI plasma region.
The electric field acting on it approaches zero with time.
Particle  2  (Fig. 14(b)) not only falls into the region $ x> 0 $, but also gradually moves in the positive direction inside the region of the EI plasma.
The coordinate and velocity of the particle undergo oscillations.
Obviously, this is due to the fact that this particle is trapped and is in a potential well.
The same oscillations also appear on the time dependence of the electric field at the point where the particle is located.
And finally, particle 3 (Fig. 14(c)) with an initial coordinate close to the boundary $ x = 0 $ demonstrates the particle acceleration process.
The velocity of this particle monotonously and quite rapidly increases from zero to $3.06$, after which it remains at a constant value.
The acceleration of the negative ion ceases after it leaves the region of a nonzero electric field.
The electric field acting on this particle, first increases in magnitude, and then falls.

Note that similar acceleration and trapping of negative ions also occur at lower absolute values of the initial amplitude.
The phenomenon is still observed at $ \varphi_ {m1}=-0.2 $, but at  $ \varphi_ {m1}\le-0.15 $ there are no accelerated and trapped negative ions.

\section{Concluding remarks}

In this paper, we studied the passage of solitary waves through the boundary between an EI plasma and a  NI plasma.
It was found that during the passage of a solitary wave from an  EI plasma to a NI plasma, a disturbance  arises in the latter, from which a solitary wave and a chain of oscillations form over time.
In almost the same way, a compressive solitary wave passes from a NI plasma to an EI plasma.
Note that a similar scenario is also observed during the evolution of a limited in size  compressive disturbance in a plasma, which also decays over time into one or several solitary waves and oscillations \cite {Karpman, Okutsu}.

In such passages, the amplitude of a solitary wave in an EI plasma, as a rule, turns out to be larger than its amplitude in a NI plasma.
An exception is the case when the initial amplitude of a compressive solitary wave in the  NI plasma is close to or exeeds the critical amplitude  in the EI plasma.
Since the critical amplitude cannot be exceeded, the amplitude of the resulting solitary wave in the EI plasma should be set at a level below the critical amplitude.
A decrease in the amplitude of a solitary wave means that its energy decreases.
A decrease in energy can occur as a result of the breaking of a solitary wave of large amplitude when it enters an EI plasma.
The breaking occurs due to the fact that some of the positive ions acquire velocities that exceed the propagation velocity of the solitary wave in the EI plasma.
Such fast ions overtake the solitary wave peak and find themselves in the region of an accelerating electric field, where they acquire even greater velocities.
Then, the accelerated ions leave the solitary wave region and continue to move along the unperturbed plasma region.
The acceleration of ions occurs due to the energy of the solitary wave.
As a result, the amplitude of the resulting solitary wave decreases to a value below the critical value.

A similar phenomenon of energy release by solitary waves was also observed during head-on collisions of solitary waves of large amplitudes \cite {PhysCom, PhysPlaRep, EurJD}.
Such collisions slow down the colliding solitary waves and then their propagation with previous velocities becomes impossible.
The transfer of excess energy to ions leads to a change in the amplitude of the solitary wave in accordance with the changed velocity of its propagation.
The solitary wave after the collision remains a solitary wave, although with a reduced amplitude.
It is known that solitary waves of not very large amplitudes, solitons, preserve their amplitudes and remain solitons in  mutual collisions.
The processes occurring both in the passage of solitary waves of large amplitudes from a NI plasma to an EI plasma, and in collisions of solitary waves of large amplitudes, demonstrate the ability of solitary waves to remain solitary waves, although somewhat changed, even under adverse external influences.
 
 Another scenario is observed when a rarefactive solitary wave enters from a NI plasma into an EI plasma.              
A rarefactive solitary wave cannot exist in an EI plasma.
Therefore, a disturbance in the form of a chain of oscillations arises in the EI plasma.
Negative ions, which were initially close to the boundary, are shifted to the region of the EI plasma.                                                         
Some of these ions fall in the region of the negative electric field and are accelerated by this field, and then, having acquired rather high velocities, they move along the unperturbed region of the EI plasma.
In addition to such accelerated ions, a group of negative ions trapped by oscillations is  observed in numerical experiments.
 Oscillations propagate at subsonic velocity and decay over time.
Therefore, the trapped ions gradually turn into free negative ions brought into the EI plasma.
Here the situation is significantly different from the case when negative ions are trapped  by a compressive solitary wave \cite {Trap}.   
In the latter case, a solitary wave propagates without changes and can transfer a sufficiently large group of negative ions over a long distance.

\end{document}